\documentclass[apjl]{emulateapj}

\slugcomment{Accepted for publication in ApJ Lett}

\usepackage{natbib}
\usepackage{graphicx}

\shorttitle{Outer stellar halo of NGC 5128}
\shortauthors{Rejkuba et al.}

\begin{document}


\title{Tracing the outer halo in a giant elliptical to 25 R$_{\mathrm{eff}}$}


\author{M. Rejkuba\altaffilmark{1,2},  W.~E.~Harris\altaffilmark{3}, L. Greggio\altaffilmark{4}, G.~L.~H.~Harris\altaffilmark{5}, H. Jerjen\altaffilmark{6}, O.~A. Gonzalez\altaffilmark{7}}
\affil{$^1$ESO, Karl-Schwarzschild-Strasse 2, D-85748 Garching, Germany}
\affil{$^2$Excellence Cluster Universe, Boltzmannstr.\ 2, D-85748, Garching, Germany}
\affil{$^3$Department of Physics and Astronomy, McMaster University, Hamilton ON L8S 4M1, Canada}
\affil{$^4$INAF, Osservatorio Astronomico di Padova, Vicolo dell'Osservatorio 5, Padova, Italy}
\affil{$^5$Department of Physics and Astronomy, University of Waterloo, Waterloo ON N2L 3G1, Canada}
\affil{$^6$Research School of Astronomy and Astrophysics, Australian National University, Cotter Rd., Weston ACT 2611, Australia}
\affil{$^7$European Southern Observatory, Ave. Alonso de Cordova 3107, 19001 Casilla Vitacura, Santiago, Chile}
\email{mrejkuba@eso.org}

%
%




\begin{abstract}
We have used the ACS and WFC3 cameras on board HST to resolve stars in the
halo of the nearest giant elliptical (gE) galaxy NGC 5128 out to a projected 
distance of 140 kpc (25 effective radii, R$_{\mathrm{eff}}$) along the major axis 
and 90 kpc (16 R$_{\mathrm{eff}}$) along the minor
axis. This dataset provides an unprecedented radial coverage of the stellar
halo properties in any gE galaxy.\\
Color-magnitude diagrams clearly reveal the presence of the red giant
branch stars belonging to the halo of NGC 5128, even in our most distant
fields. The star counts demonstrate increasing flattening of the outer halo, which is 
elongated along the major axis of the galaxy. The V-I colors of the red giants enable us to 
measure the metallicity distribution in each field and so map the 
gradient out to $\sim 16$ R$_{\mathrm{eff}}$ from the galaxy center along the major axis.
A median metallicity is obtained even for the outermost fields along both axes.
We observe a smooth transition from a metal-rich ([M/H]$\sim 0.0$) inner galaxy to 
lower metallicity in the outer halo, with the metallicity gradient slope along the major axis of 
$\Delta$[M/H]/$\Delta R$ $\simeq -0.0054 \pm 0.0006$ dex~kpc$^{-1}$. 
In the outer halo, beyond $\sim 10$ R$_{\mathrm{eff}}$, the number density profile 
follows a power law, but also significant field-to-field metallicity 
and star count variations are detected. 
The metal-rich component dominates in all observed fields, and the median metallicity is 
 [M/H]$>-1$~dex in all fields.

\end{abstract}



\section{Introduction}

Because dynamical timescales in the outer halos of large galaxies are long, the properties of stars in the remote halo regions, such as their metallicity and surface density distributions and gradients, provide important constraints for galaxy formation and hierarchical assembly models \citep{johnston+08,font+11,naab+ostriker09,cooper+13}.  However, the extremely faint surface brightness of the halo makes the raw observations challenging.  Integrated-light studies need to probe surface brightness levels of $\mu \sim 30-32$ mag~arcsec$^{-2}$, predicted in simulations for outer halos \citep[e.g.][]{johnston+08} where most of the sub-structure tracing the hierarchical assembly history of the galaxies is expected. Such faint limits are out of reach for  integrated light studies with
most modern large telescopes \citep{jablonka+10,abraham+vandokkum14}, though attempts have been made to derive \emph{mean} properties by  stacking many similar galaxies \citep{zibetti+04,tal+vandokkum11}.

A powerful alternative to integrated light studies is resolved photometry of individual stars, a technique that can be applied for galaxies at distances $\lesssim 20$ Mpc.  Direct stellar photometry is capable of revealing not just mean halo properties but the \emph{distribution functions} of metallicity and age.

Only a few giant elliptical (gE) galaxies are close enough to allow detection and 
study of their outer halo
regions. \citet{mihos+13} traced the halo color gradient of the Virgo cluster 
gE M49 out to $\sim 100$~kpc (7~R$_{\mathrm{eff}}$), where $\mu_V \sim 28$~mag arcsec$^{-2}$. 
The bluer color of the outer halo and the steepening of the color gradient in the M49 halo, if 
interpreted as  purely due to metallicity, argues for the presence of a dominant 
metal-poor population 
with [Fe/H]$<-1$. The detection of this ``classical" metal-poor halo is in contrast with
the findings in the Centaurus  A (Cen A) 
group central giant NGC 5128, where both a high mean metallicity and no strong 
metallicity gradient 
have been seen out to a similar scaled distance of 6.5~R$_{\mathrm{eff}}$ \citep{rejkuba+05}. 
In the Leo group gE NGC~3379 the transition to a classically metal-poor outer halo was 
found at $\sim 10-12$~R$_{\mathrm{eff}}$ \citep{harris+07b}.
These few galaxies and very limited spatial coverage (with only 1 or few pointings per galaxy) 
preclude drawing general conclusions and demand a wider range of observations.

NGC~5128 (often referred to as Cen A, after its radio source name) is a particularly important target. 
A mere 3.8~Mpc distant \citep{harris+10}, it is by far the closest and easiest to observe gE  galaxy.
Its old-halo red-giant stars can be studied readily with Hubble Space Telescope (HST) imaging 
\citep[e.g.\ ][]{soria+96, marleau+00, harris+99} or under excellent seeing conditions from 
the ground \citep{rejkuba+01, rejkuba+03, crnojevic+13}. 
HST-based photometry, reaching at least 1-1.5 mag below the tip of the RGB, has allowed the metallicity 
distribution function (MDF) to be investigated in detail in four fields at projected galactocentric 
distances from  8 to 38 kpc \citep{harris+99,harris+harris00,harris+harris02,rejkuba+05}. 
The outer-halo stars were traced out to $\sim 85$~kpc with VIMOS at the ESO Very Large 
Telescope \citep{crnojevic+13} and found to have high median metallicity even 
at $\sim 15 R_{\mathrm{eff}}$\footnote{\citet{crnojevic+13} used R$_{\mathrm{eff}} =330$ arcsec=6.1~kpc, while we adopt  305 arcsec = 5.6~kpc from \citet{dufour+79}.}. 
This VIMOS study also provides a first hint that the outer halo becomes more elongated and has a surface brightness profile shallower than the  r$^{1/4}$ law. 
However, high contamination by both foreground stars and unresolved background
galaxies prevented stronger conclusions.  Deeper and higher-resolution HST observations can go 
much further.

In this work we extend our previous HST studies of the NGC~5128 halo into the extreme outer 
parts of the halo up to 140~kpc or 25~R$_{\mathrm{eff}}$, regions never observed before. With 
these new data we address the questions: (i) how far does the halo extend? (ii) is there  
a genuine metal-poor halo in NGC~5128, similar to that found in the outer regions of the 
Milky Way, M31, NGC~3379 and M49 \citep{carollo+07, chapman+06, harris+07b,mihos+13}? 
(iii) and if so, at which radius does the metal-poor population start to dominate?

\section{Observations and data reduction}

Five new fields were observed with the HST during Cycle 20 (Cy20; GO programme 12964) with 
the Wide Field Camera 3 (WFC3) in UVIS mode as primary and Advanced Camera for Surveys 
(ACS) in Wide Field Camera mode as the secondary (parallel) instrument. For both instruments 
imaging was done with the F606W and F814W, the same filters as used in all our previous 
studies.  Each field received three dithered exposures adding to one orbit per filter. The total 
exposure time for WFC3 was 2376 sec (F814W) and 2496 sec (F606W), while the ACS images got 
2137 and 2270 sec in the same two bands. 

\begin{figure}
\includegraphics[angle=0,width=9cm]{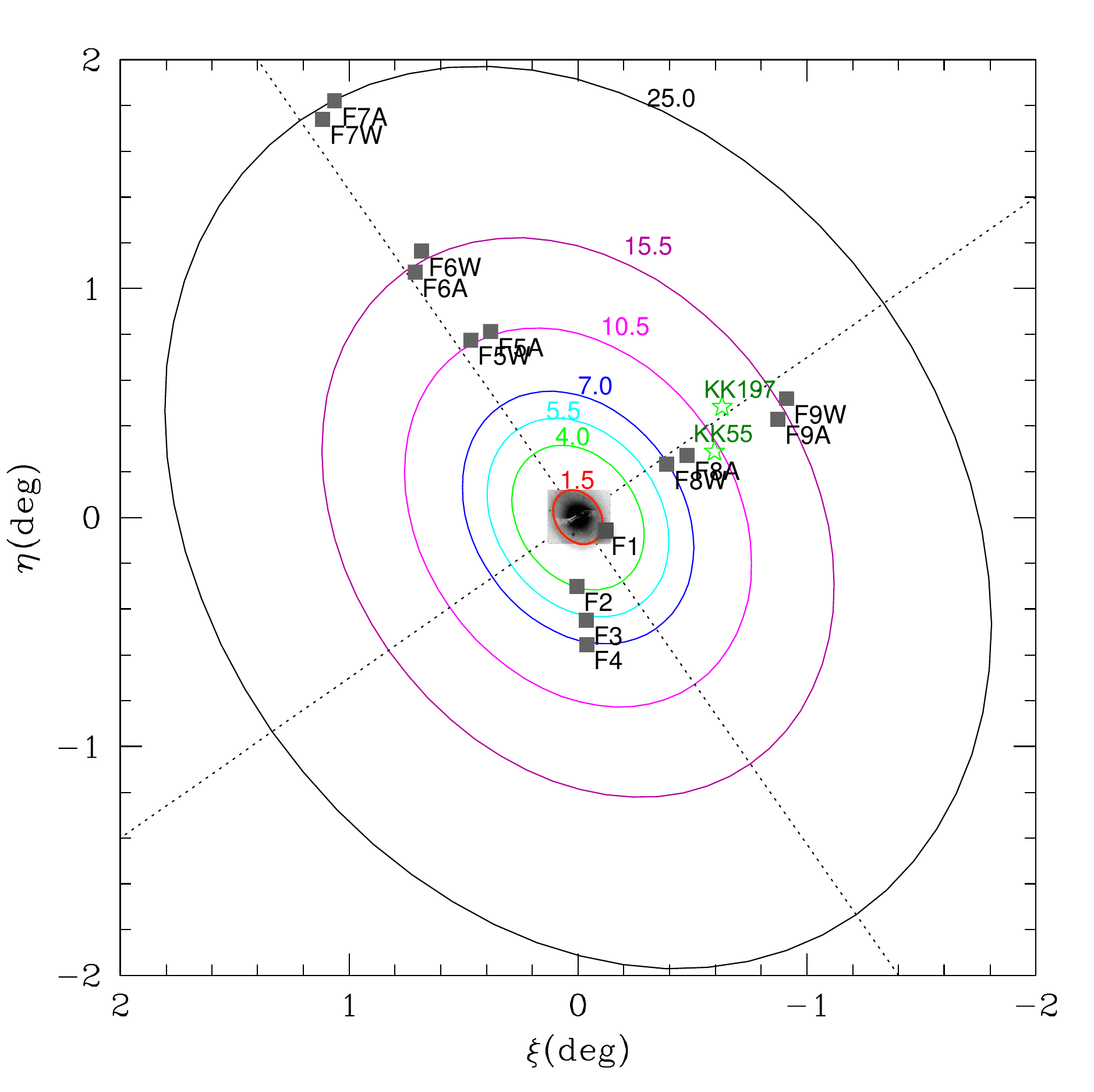}
\caption{Distribution of the HST observations in the halo of NGC~5128 relative to the center of the galaxy projected on the plane of the sky. The elliptical contours have an axis ratio of 0.77 and PA=35$^\circ$ and are plotted for 1.5, 4.0, 5.5, 7.0 10.5, 15.5 and 25 
R$_{\mathrm{eff}}$ distance, which are representative for our present and past HST pointings. The two green stars are dwarf spheroidal galaxies in the Cen A group, KK197 and KKs55. \label{fig:fieldsrel}}
\end{figure}

Figure~\ref{fig:fieldsrel} shows the locations of our fields relative to the center of NGC~5128 
projected on the plane of the sky. Fields F1-F4 are described in \citet{harris+99}, 
\citet{harris+harris00,harris+harris02}, and \citet{rejkuba+05}, while our new Cy20 fields F5-F9 
are marked with suffix A (ACS) or W (WFC3). Following \citet{dufour+79}, the contour lines 
shown are for an axis ratio 0.77, with the major axis along a position angle of 35$^\circ$.  The 
effective radius is $R_{\mathrm{eff}} = 305''$ corresponding to 5.6 kpc at the 3.8~Mpc distance 
\citep{harris+10}. Our new fields F5, F6, and F7 are located along the major axis $\sim 60$, 90 and 140 kpc North-East from the center of NGC 5128, respectively.   The fields F8 and F9 along the North-West minor axis lie on the same ellipses as F4 and F6, which are located respectively at $\sim 40$ ($\sim 7$~R$_{\mathrm{eff}}$) and 90 kpc ($\sim 16$  R $_{\mathrm{eff}}$) close to and along the major axis. 

The pipeline processed images were downloaded from the HST archive, and pixel-based charge transfer corrections 
\citep{anderson+bedin10}  were applied with the Fortran 
code\footnote{http://www.stsci.edu/hst/wfc3/tools/cte\_tools} from J.\ Anderson \citep{ubeda+anderson12}.  PSF-fitting photometry was completed with the DAOPHOT suite of programmes \citep{stetson87}. Our images are completely uncrowded, so in principle aperture photometry would be accurate, but PSF/ALLSTAR fitting was preferred to allow us  using the DAOPHOT shape parameters ($\chi$, $sharp$) to cull nonstellar objects (faint, small background galaxies) as far as possible. 
Stars detected were retained if (i) their positions on both filters matched to within 1 px, (ii) $\chi< 1.5$, ABS($sharp$)$<0.4$, and  (iii) measurement uncertainties were smaller than 0.5 mag.  Photometric calibration in the VEGAMAG system followed the prescription in \citet{sirianni+05}, with the latest zero points recommended on the instrument web pages\footnote{http://www.stsci.edu/hst/wfc3/phot\_zp\_lbn; http://www.stsci.edu/hst/acs/analysis/zeropoints}.  Final transformation to the ground-based $VI$ magnitude scale follows \citet{saha+11}.

\section{Results}

\subsection{Color Magnitude diagrams} 
\label{sec:cmds}

\begin{figure*}
\includegraphics[angle=270,width=17cm]{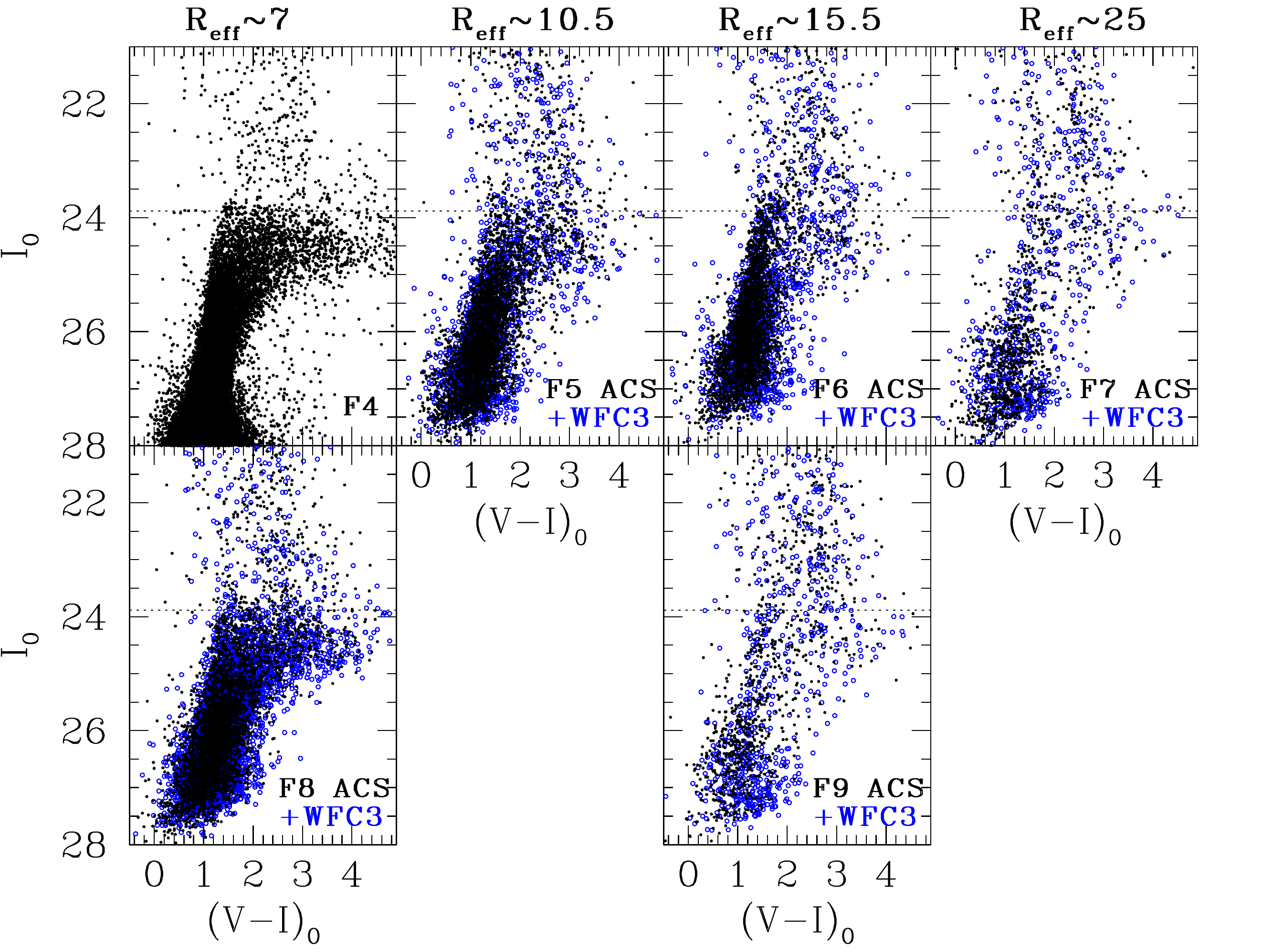}
\caption{Color-magnitude diagrams of the stellar sources detected in the fields in the halo of NGC~5128. F4 is the ACS field previously obtained by \citet{rejkuba+05}, while the other 5 panels show the new CMDs observed with WFC3 (blue open circles) and ACS (black dots) during our Cy20 HST program. 
The dotted horizontal line indicates the tip of the RGB. \label{fig:cmds}}
\end{figure*}

In Figure~\ref{fig:cmds} we show the color-magnitude diagrams (CMDs)  for the new fields, in comparison with the deep CMD in F4 from Cycle 11 \citep{rejkuba+05}. The top row displays the CMDs in fields along the major axis (F5-F7), while the bottom row shows the CMDs for the two fields along the minor axis (F8-F9). The CMDs for fields along the minor axis fields are plotted immediately below the 
corresponding CMD of the major axis field that is on the same ellipse (c.f.\ Fig.\ref{fig:fieldsrel}). The magnitudes and colors are extinction-corrected for each field individually following the \citet{schlafly+finkbeiner11} recalibration of the Galactic extinction maps of 
\citet{schlegel+98}, assuming the \citet{fitzpatrick99} reddening law with $R_V=3.1$.
The dotted horizontal line marks the expected magnitude of the tip of the RGB at $I_0=23.88$ \citep{harris+10}. 

The CMDs in Fig.~\ref{fig:cmds} show a very similar, broad upper RGB in fields F4 and F8; F5 is almost as broad in color.  F6, however, has a narrower upper RGB, in particular in the ACS pointings (black dots), suggesting the beginning of a transition to a more metal-poor-dominated population.  In F7 and F9, the RGB is quite sparsely populated, but the color distribution still resembles that of F6. 

\begin{figure*}
\includegraphics[angle=0,width=8.5cm]{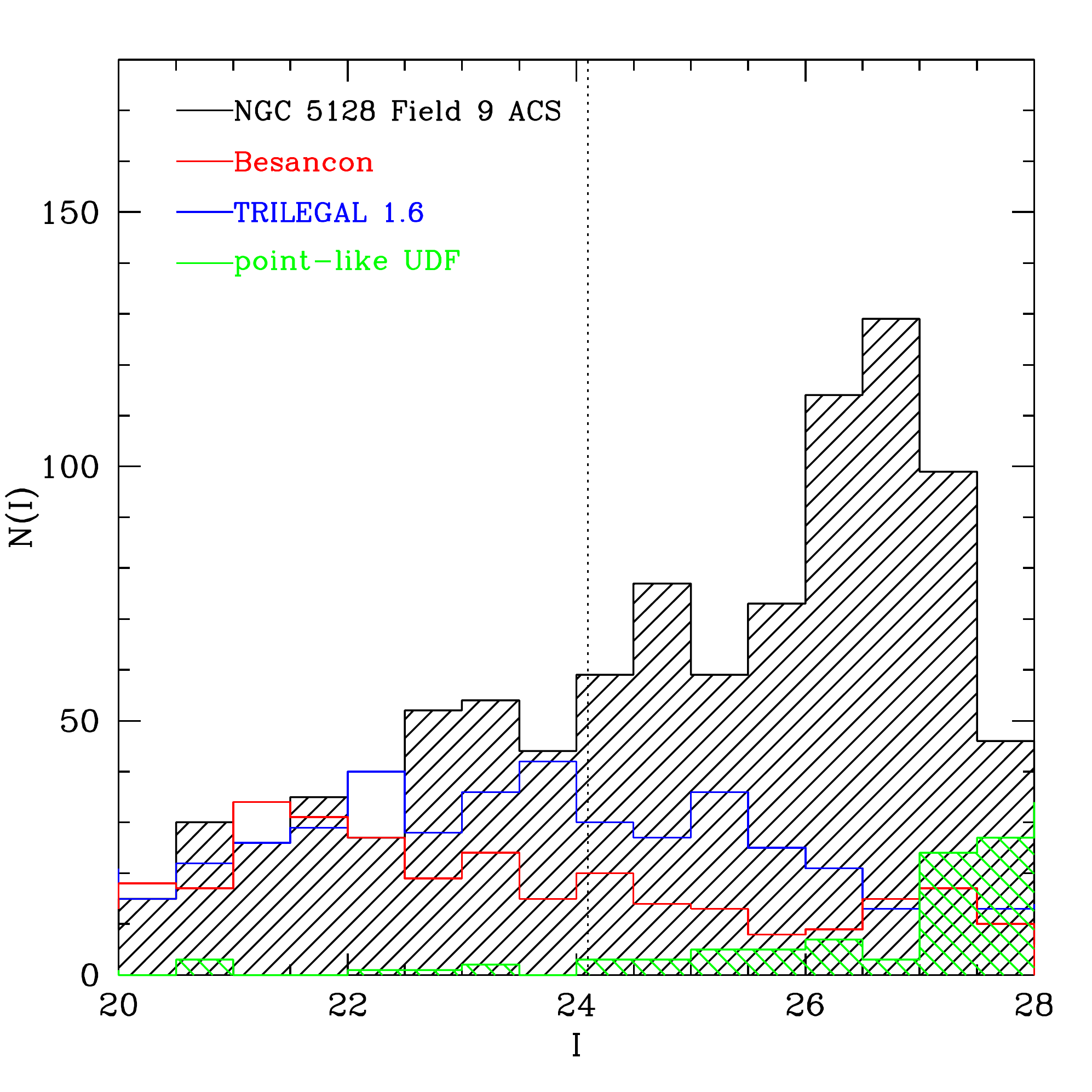}
\includegraphics[angle=0,width=8.5cm]{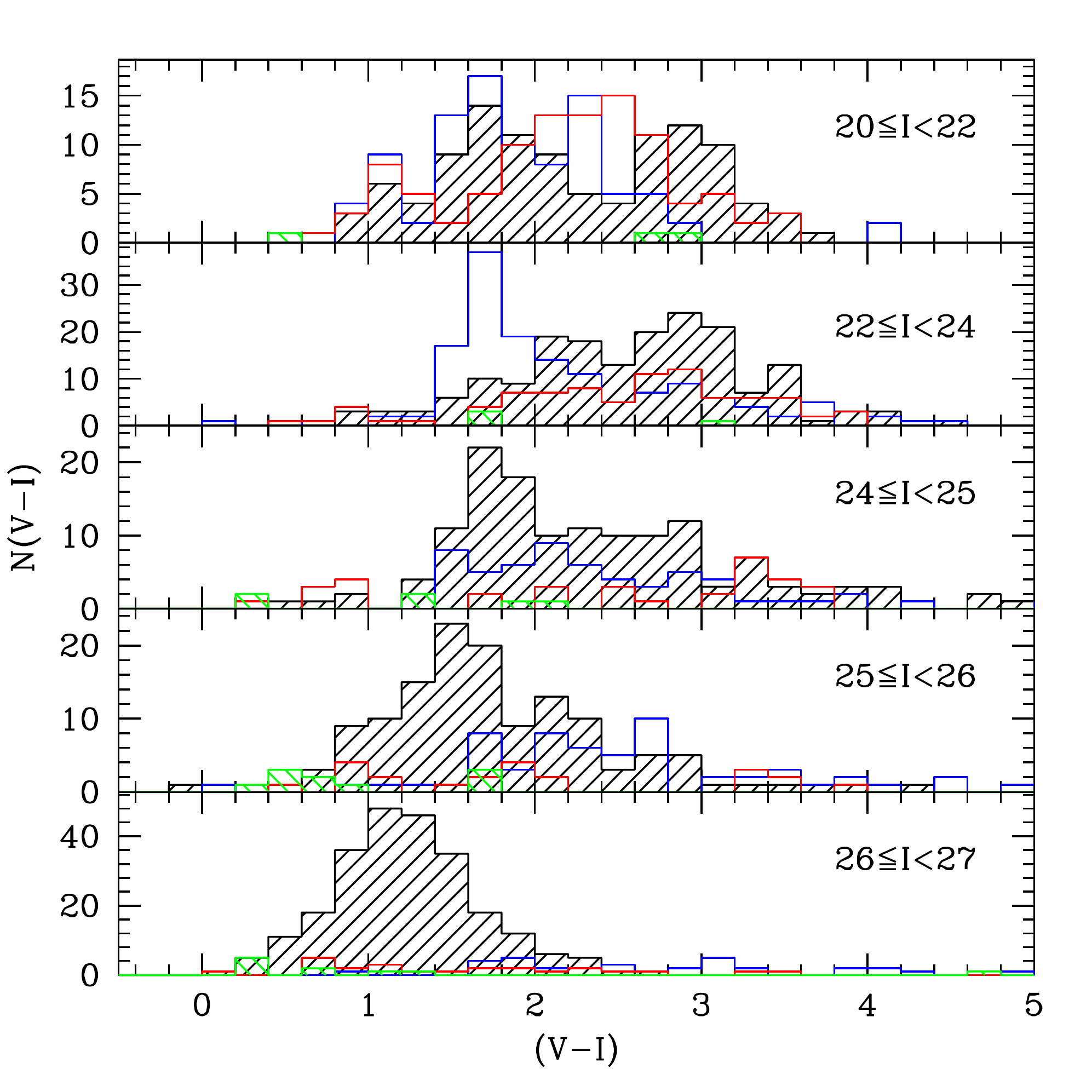}
\caption{Left: The observed I-band luminosity function in the F9 ACS field (black hashed histogram) compared with the simulated MW foreground luminosity function from the TRILEGAL v1.6 (blue), and Besan\c{c}on model (red) of the Galaxy, and with the UDF point-like sources (green hashed histogram). 
The vertical dotted line at I=24.1 indicates the position of the RGB tip magnitude in NGC~5128. 
Right: The color distribution of the F9 ACS field (black hashed histogram) in different magnitude bins compared with the simulated foreground stars from TRILEGAL v1.6 (blue), Besan\c{c}on (red), and UDF point-like sources (green histogram). Simulated and observed distributions include reddening.
\label{fig:contamination}}
\end{figure*}

Most of the objects brighter than $I_0=23.88$ (i.e. clearly above the RGB tip) are foreground Milky Way 
(MW) stars, but a few bright AGB stars are expected in NGC~5128 halo, based on the study of 
Long Period Variables \citep{rejkuba+03}, and 38~kpc field star formation history
\citep{rejkuba+11}. The camera areas ratio (ACS/WFC3) is 1.55, and 
we thus expect the same ratio for the number of bright stars, if they all belong to the MW foreground. 
This is indeed observed in F7 and F9, but there are  $1.8 \times$ fewer stars in F5, F6 and F8 WFC3 
pointings compared to ACS. This might be due to a small variation of contribution from bright AGB stars in NGC~5128 or
sub-structure in the MW foreground, and will be investigated in a follow-up study.
At levels fainter than the tip of the RGB
($I_0 > 23.88$),  foreground stars are also present, and at the faintest magnitudes ($I \gtrsim 27$) 
compact high-redshift galaxies may also add to the contamination. 

To estimate the foreground contamination quantitatively we use the TRILEGAL \citep{girardi+05} 
and Besan\c{c}on \citep{robin+03} MW models, while the contamination by background 
unresolved galaxies can be evaluated from the Hubble Ultra Deep Field (UDF) 
observations\footnote{http://heasarc.gsfc.nasa.gov/W3Browse/hst/hubbleudf.html}. 
We transformed the UDF F775W  AB measurements to F814W  on Vegamag photometric system adopting the relation 
from \citet{sirianni+05}.
An example of the model field components is shown in Figure \ref{fig:contamination} for F9.  
The model starcounts above the RGB tip reveal significant differences between TRILEGAL and 
Besan\c{c}on. Though the investigation of these differences is beyond the scope of this work, we 
decided to adopt TRILEGAL (version 1.6), because it had about 30\% more stars with respect 
to the Besan\c{c}on simulation, and it better matched the bright end of the observed luminosity 
function (Fig.~\ref{fig:contamination}, left), and the color distribution of the bright foreground stars 
with $20\leq I < 22$ (Fig.~\ref{fig:contamination}, right). 

The model field population from TRILEGAL run at each field location, normalized to the ACS/WFC and WFC3 areas,
was used to statistically clean the observed CMDs in all five fields.  This step removed almost all 
stars brighter than the RGB tip, except stars redder than $V-I>2.8$, which are missing in the 
TRILEGAL simulations.   The simulations include reddening, but were not corrected for incompleteness and therefore it is 
likely that the faintest stars are oversubtracted. However, since we are interested in the upper  $\sim 1.6$
magnitude below the RGB tip, from which we measure the metallicity distributions, and the 
photometry is highly complete at that level, any such corrections will have no significant effect on the MDFs.
Remaining potential contaminants from unresolved galaxies 
are mostly fainter and bluer than the stars used to derive MDFs
(Fig.~\ref{fig:contamination}, right). This, combined with their low total number shows that the 
unresolved galaxies do not affect our metallicity distribution derivation.

The fact that the RGB is still detectable in F7 at 140~kpc ($25 R_{\mathrm{eff}}$) is noteworthy by itself; our original expectation in designing the observations was that it would act as a ``control field'' of almost pure field population based on outward extrapolation of the inner-halo $r^{1/4}$ density profile.  In short, we have not yet found the limits of this giant galaxy's halo. It is also striking that the CMD in the F9 minor axis field, lying on the same ellipse as F6 (the major axis field at $16~R_{\mathrm{eff}}$), appears similar to F7 at the much larger distance of $25~R_{\mathrm{eff}}$  along the \emph{major} axis field; if the 
isophotal contours kept the same ellipticity with radius, then the level of star counts in F9 
should be the same as in F6, but instead it is considerably lower, and similar to that in F7 (see Sec.~\ref{sec:gradient}). 
This evidence suggests that either the outer halo becomes increasingly elongated, as already 
suspected from the VIMOS observations by \citet{crnojevic+13}; or that some sort of very extended 
tidal debris plume lies along the major axis, boosting the level of starcounts over the normal smooth 
halo component.

\subsection{Halo Metallicity and Number Density Profiles}
\label{sec:gradient}

\begin{figure*}
\includegraphics[angle=270,width=17cm]{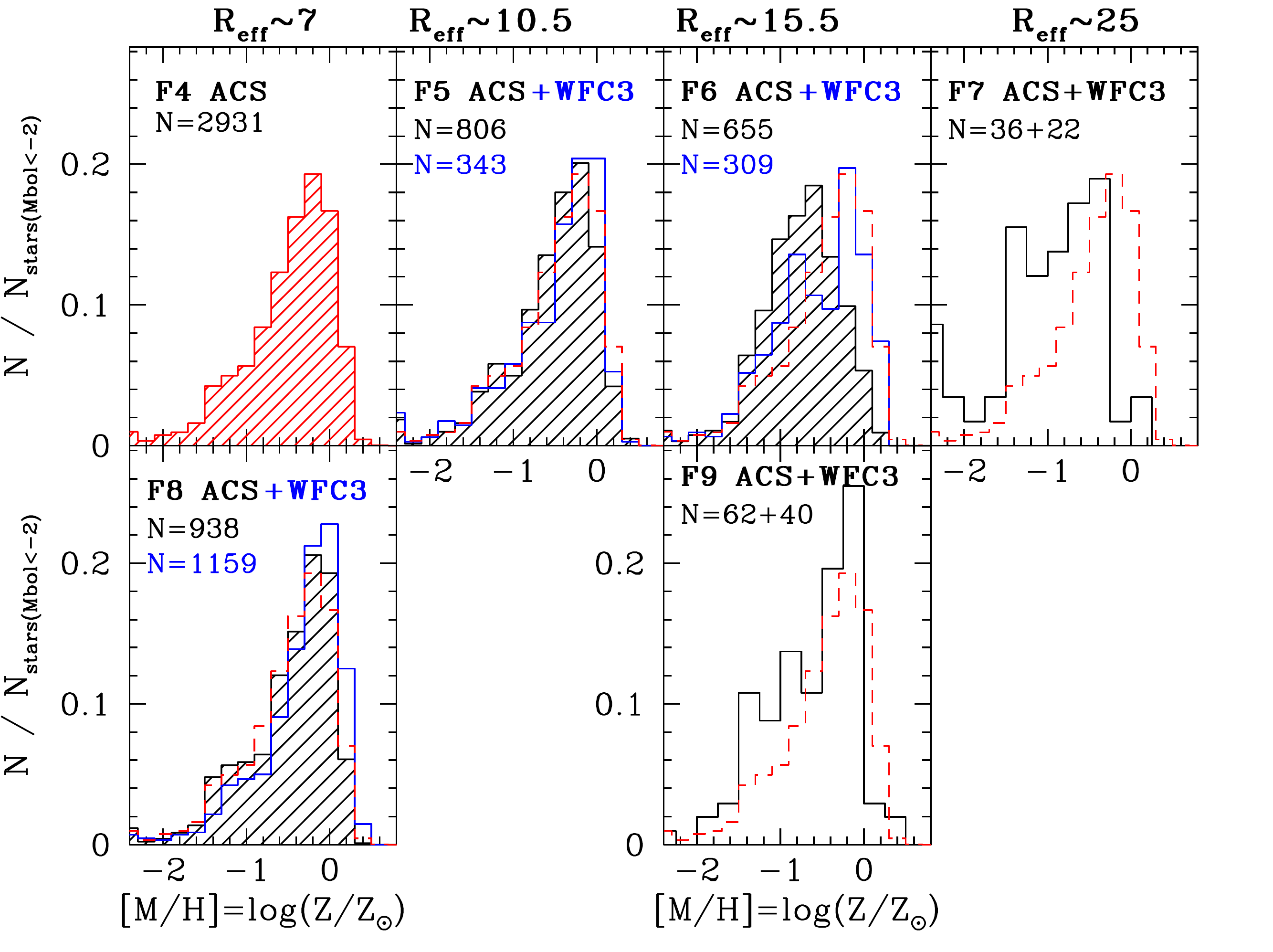}
\caption{The metallicity distribution functions derived from 
contamination-cleaned CMDs. The red (dashed) histogram in each panel is the F4 MDF. The blue is for the 
WFC3 data and black for ACS. For F7 and F9 the ACS and WFC3 measurements are combined (solid black histogram). All histograms are normalized by the total number of stars shown in each panel. \label{fig:MDFs}}
\end{figure*}

Our analysis of the very deep F4 photometry, which reached the red clump helium burning stars, 
quantitatively demonstrated that 70-80\% of the stars have classically ``old'' ages in excess of $\sim 
10$~Gyr \citep{rejkuba+11}. The colors of the old RGB stars can therefore be used also for other halo fields to derive the 
MDF without complications from the age/metallicity degeneracy.  
Conversion of $(V-I)_0$ to heavy-element abundance $Z$  for stars with $-3.6<M_{bol}<-2$ 
was done by interpolation in an isochrone grid as described in \citet{harris+harris00}, with the 
$\alpha$-enhanced Teramo isochrones \citep{pietrinferni+06} as done for F4 \citep{rejkuba+11}. 
The MDFs for the inner  fields 1-3 (observed with WFPC2) have also been re-derived with this 
newer isochrone set. 
The absolute metallicity scale coming from the use of the new set of isochrones 
results in slightly higher average metallicity  (by about 0.2 dex) than reported 
in \citet{rejkuba+05}. However, all metallicities derived in this work are on the same scale and 
thus a \emph{relative} comparison between the
distributions of different fields is robust and can be used to investigate the halo metallicity 
gradient.

\begin{figure*}
\includegraphics[angle=0,width=8.5cm]{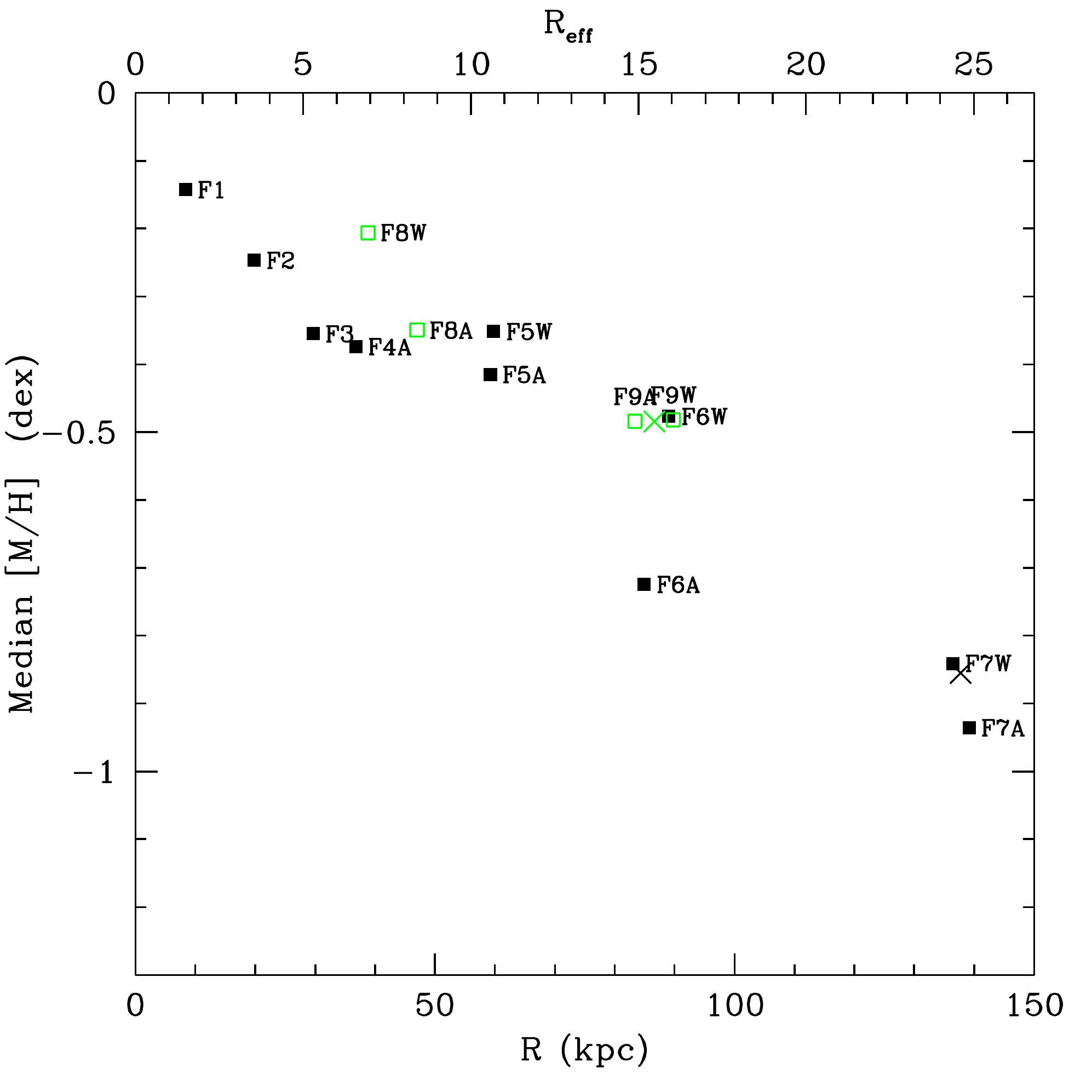}
\includegraphics[angle=0,width=8.5cm]{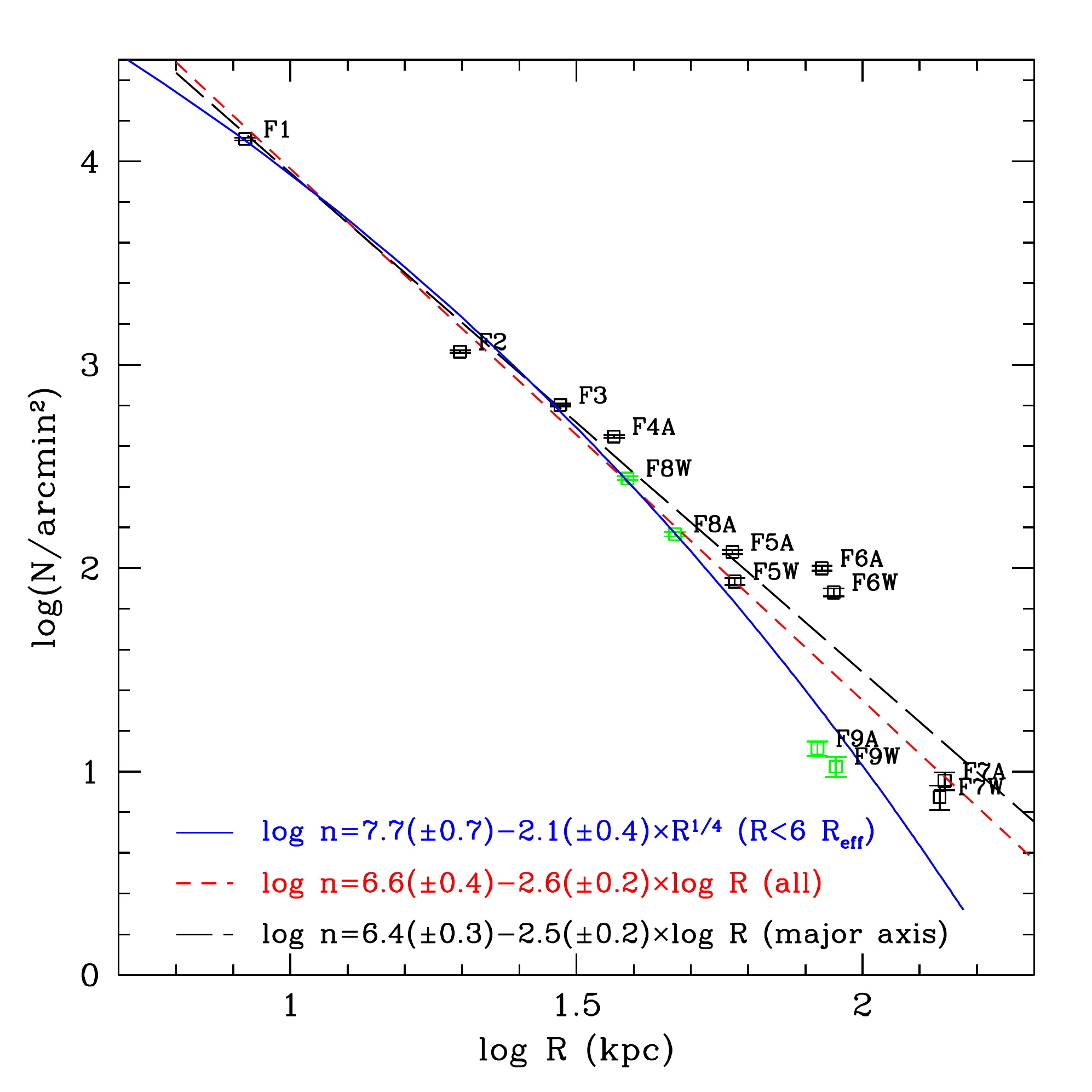}
\caption{Median metallicity (left) and density profile of RGB stars with $24\leq I_0 <26$ (right) as functions of the projected distance from the center of NGC 5128 along the major axis. 
Open green squares indicate the minor axis fields, reported at their projected distance multiplied by the inverse axis ratio (a/b), thus referred to the ellipse to which they belong (Fig.~\ref{fig:fieldsrel}). 
In the left panel, crosses show the results obtained by combining the ACS and WFC3 data for F7 and F9 fields.
In the right panel we plot different fits to the density profile: a R$^{1/4}$ law applied to the inner 6 R$_{\mathrm{eff}}$ (solid blue), a power law applied to all (short dashed red) and only major axis (long dashed black) fields.
\label{fig:gradient}}
\end{figure*}

Figure~\ref{fig:MDFs} shows the new MDFs for all our fields F4-F9, derived from the 
foreground cleaned CMDs. The metallicity distributions are wide in 
all the panels, but in F7 and F9 they are poorly sampled because of small number statistics.
We thus show combined WFC3 and ACS MDFs for F7 and F9. 
For comparison in all panels we overplot the MDF measured from ACS F4 dataset.

The left panel of Figure~\ref{fig:gradient} shows the \emph{median} metallicity as a function of projected distance 
from the center of the galaxy along the major axis. The green open symbols denote minor axis fields (F8 and F9), for 
which the distance is scaled to the major axis by multiplying with the inverse axis ratio 
(a/b=1/0.77). A clear metallicity gradient exists, with slope 
$\Delta$[M/H]/$\Delta R$ $\simeq -0.0054 \pm 0.0006$ dex per kpc, or 
$\Delta$[M/H]/$\Delta R_{\mathrm{eff}}$ $\simeq -0.030 \pm 0.003$ dex per R$_{\mathrm{eff}}$.  
Remarkably, however, even in the most remote parts of the halo we have sampled, the median metallicity has not 
fallen below [M/H] $= -1$~dex.  

Field-to-field scatter also shows up, such as much lower [M/H] in the ACS vs.\ WFC3 pointing of F6.   
For F7 and F9 fields we plot in addition to individual ACS and WFC3, also a median [M/H] from both 
cameras together (crosses). 
Bootstrap simulations with 5000 random extractions of smaller  samples representative for F7 and F9 MDFs, show that 
the median metallicity can be derived with $\lesssim 0.1$~dex accuracy for samples 
with $\geq 40$ stars, while the uncertainty increases to $\sim 0.2$~dex for samples 
of 20--40 stars. These simulations demonstrate that the metallicity difference between F7 and F9 is 
significant in spite of low statistics.

The difference between the F7 and F9 mean metallicity is even more remarkable considering 
that these two fields have similar
stellar density. The right panel of Figure~\ref{fig:gradient} shows the number density of RGB stars with 
$24\leq I_0<26$ (selected from foreground cleaned CMDs) as a function of the projected distance along the major axis.
Previously mentioned striking difference between the minor and major axis fields located at the 
same projected major axis distance
(F6 and F9) is even better appreciated here. The minor axis fields density profile follows the 
extrapolation of $r^{1/4}$ profile, while a power law provides a better fit along the major axis 
beyond R$_{\mathrm{eff}}\sim 10$, where the stellar density exceeds
the shallower $r^{1/4}$ profile.
Additional field-to-field variations seen in metallicity are also observed in density distribution gradient. 
We may, perhaps, be seeing the record of incompletely mixed debris from hierarchical buildup of 
the halo.

\section{Discussion and Conclusions}

The new HST observations presented here trace the halo of an early-type giant galaxy to 
unprecedentedly large distances ($25 R_{\mathrm{eff}}$). It is interesting to put this distance
in a context of cosmological "virial radius" $R_{200}$, at which the mean density equals 
200$\times$ the cosmological critical density. Taking the total mass of 
$(9.2 \pm 0.3) \times 10^{12}$ M$_\odot$ for the entire Cen A group 
\citep[][using dynamics of all the satellites]{woodley06}, would imply
that the Cen A group virial radius is $R_{200}=400$ kpc, and our outermost 
field at 140 kpc corresponds to 1/3 of the virial radius.

Comparison of the RGB star counts between the major and minor axes suggests increasing 
elongation of the halo  beyond $\sim 10 R_{\mathrm{eff}}$.  While sub-structure is expected to be 
present in the outer halo, the significantly lower density of stars at 16~R$_{\mathrm{eff}}$ 
along the 
minor vs.\ major axis is also confirmed by the detection of stars at much larger distance along the 
major axis. This result furthermore confirms a tentative detection of the increasing 
ellipticity in the outer halo by \citet{crnojevic+13}. 
This result is also interesting in comparison with the average ellipticity of stellar 
halos of early-type galaxies from \citet{tal+vandokkum11}, which is based on a very deep stack 
of more than 42,000 SDSS images of luminous red galaxies. These authors found an increasing 
ellipticity with radius from $\epsilon=0.25$ at 10 kpc to $\epsilon=0.30$ at 100~kpc. The ellipticity 
in the halo of NGC~5128 appears to increase even more, although our so-far sparse sampling of 
the halo is only suggestive. 

\citet{tal+vandokkum11} furthermore found that the surface brightness profile of their average stellar halo follows a Sersic model out to 100~kpc, but an excess of light beyond that is present in r, i and z-band stacks. They discuss whether this excess light is part of the galaxy haloes or it traces the unresolved intragroup or intracluster light.  With our measurement of the NGC 5128 metallicity gradient and starcounts we find no clear transition to intragroup light beyond 90~kpc (beyond F6) in NGC~5128. Observations of further more distant fields along the minor and major axis would be necessary however to confirm this.  

While the inner parts of the galaxy show a smooth metallicity gradient, we observe quite strong field-to-field variations in the median metallicity of the stars in the outer regions, as expected if the halo is built up by accretion of different satellites. However,  the relatively high median metallicity is surprising, and the extended halo along the major axis might suggest that we are looking at least partly at material ejected from a long-ago accretion of a disk galaxy. This conclusion is supported by the analysis of the age 
distributions of stars \citep{rejkuba+11} and globular clusters \citep{woodley+10}, as well as the presence of 
dynamical subgroupings of the globular clusters and planetary nebulae in the halo \citep{woodley+harris11}.

\acknowledgments
Based on observations made with the NASA/ESA Hubble Space Telescope, obtained at the Space Telescope Science Institute, which is operated by the Association of Universities for Research in Astronomy, Inc., under NASA contract NAS 5-26555, within program \#12964. WEH acknowledges support from the Natural Science and Engineering Research Council of Canada. HJ acknowledges financial support from the Australian Research Council through Discovery Project grant DP120100475. We thank the referee for constructive comments.

{\it Facilities:}  \facility{HST (ACS)}, \facility{HST (WFC3)}.

\end{document}